%% file: DAC24-Egraph-Opt.tex
\documentclass[sigconf,nonacm]{acmart}

\input{acm-setting}

\begin{document}

\title{\deftitle: E-Graph Rewriting with Technology-Aware Cost Functions for Logic Synthesis}

%
%

\author{Chen Chen}
\authornote{Both authors contributed equally to this research.}
\affiliation{HKUST(GZ)}
\email{cchen099@connect.hkust-gz.edu.cn}

\author{Guangyu Hu}
\authornotemark[1]
\affiliation{HKUST}
\email{ghuae@connect.ust.hk }

\author{Dongsheng Zuo}
\affiliation{HKUST(GZ)}
\email{dzuo721@connect.hkust-gz.edu.cn}

\author{Cunxi Yu}
\affiliation{University of Maryland, College Park}
\email{cunxiyu@umd.edu}

\author{Yuzhe Ma}
\affiliation{HKUST(GZ)}
\email{yuzhema@hkust-gz.edu.cn}

\author{Hongce Zhang}
\affiliation{HKUST(GZ)}
\email{hongcezh@hkust-gz.edu.cn}

\input{sections/abstract.tex}


 
\maketitle



\input{sections/intro.tex}

\input{sections/related.tex}

\input{sections/example}
\input{sections/our-method.tex}

\input{sections/experiment.tex}
\input{sections/conclusion.tex}

\input{sections/ack.tex}

\bibliographystyle{ACM-Reference-Format}
\bibliography{refs}

\end{document}

%% file: acm-setting.tex
\usepackage{amsmath,amsfonts}
\usepackage{multirow}
\usepackage{multicol}
\usepackage{amsthm}
\usepackage{listings}
\usepackage{algorithm}
\usepackage{algpseudocode}
\usepackage{graphicx}
\usepackage{textcomp}
\usepackage{xcolor}
\usepackage{cleveref}
\usepackage{booktabs}
\usepackage{comment}
\usepackage{tabularx}
\usepackage{multirow}
\usepackage{bm}
\usepackage{url}
\usepackage{xspace}
\usepackage{pifont}
\usepackage{arydshln}
\usepackage{verbatim}
\usepackage[referable]{threeparttablex}
\usepackage{calc} 
\usepackage{color}
\usepackage{float}
\usepackage{subcaption}
\usepackage[utf8]{inputenc}
\usepackage{booktabs} 
\usepackage{tabularx} 
\usepackage{float} 
\usepackage{tikz}
\usetikzlibrary{shapes.geometric, arrows}
\tikzstyle{startstop} = [rectangle, rounded corners, minimum width=3cm, minimum height=1cm,text centered, draw=black]
\tikzstyle{io} = [trapezium, trapezium left angle=70, trapezium right angle=110, minimum width=3cm, minimum height=1cm, text centered, draw=black]
\tikzstyle{process} = [rectangle, minimum width=3cm, minimum height=1cm, text centered, draw=black]
\tikzstyle{arrow} = [thick,->,>=stealth]

\newcommand{\guangyu}[1]{\textcolor{blue}{[\textbf{guangyu}: #1]}}

\newcommand{\cchen}[1]{\textcolor{purple}{[\textbf{cchen}: #1]}}

\newcommand{\deftitle}[0]{{\texttt{E-Syn}}\xspace}
\newcommand{\deftextEsyn}[0]{\mbox{\texttt{E-Syn}}\xspace}

\def\BibTeX{{\rm B\kern-.05em{\sc i\kern-.025em b}\kern-.08em
    T\kern-.1667em\lower.7ex\hbox{E}\kern-.125emX}}

 \acmConference[DAC '24]{Design Automation Conference}{June 23--27, 2024}{San Francisco, CA}

%% file: sections/abstract.tex
\begin{abstract}

Logic synthesis plays a crucial role in the digital design flow. It has a decisive influence on the final Quality of Results (QoR) of the circuit implementations. However, existing multi-level logic optimization algorithms often employ greedy approaches with a series of local optimization steps. Each step breaks the circuit into small pieces (e.g., k-feasible cuts) and applies incremental changes to individual pieces separately. These local optimization steps could limit the exploration space and may miss opportunities for significant improvements.
To address the limitation, this paper proposes using e-graph in logic synthesis. The new workflow, named \deftextEsyn, makes use of the well-established e-graph infrastructure to efficiently perform logic rewriting. It explores a diverse set of equivalent Boolean representations while allowing technology-aware cost functions to better support delay-oriented and area-oriented logic synthesis. 
Experiments over a wide range of benchmark designs show our proposed logic optimization approach reaches a wider design space compared to the commonly used AIG-based logic synthesis flow. It achieves on average 15.29\% delay saving in delay-oriented synthesis and 6.42\% area saving for area-oriented synthesis.

\end{abstract}

%% file: sections/intro.tex
\begin{figure*}
  \centering
  \includegraphics[width=\textwidth]{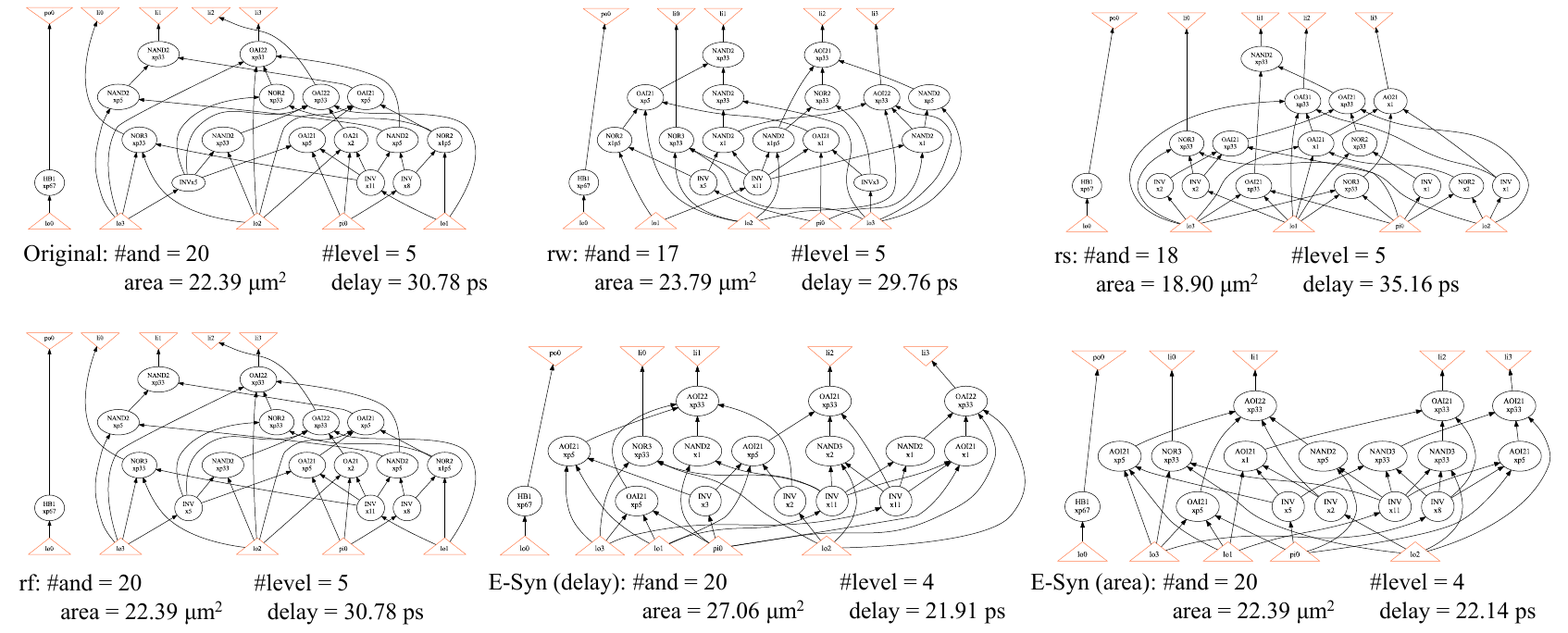}
  \caption{Netlists after technology mapping and gate sizing, using the original logic form (\texttt{original}) or after applying one of AIG rewriting (\texttt{rw}), resubstitution (\texttt{rs}), refactoring (\texttt{rf}), delay-oriented e-graph rewriting, or area-oriented e-graph rewriting}. 

  \label{fig:netlist-after-rewrite}
\end{figure*}

%
%

\section{Introduction}


Logic synthesis is a common starting point of the digital design automation flow and significantly affects the Quality of Results (QoR) such as area, timing, and power consumption.
The modern logic synthesis flow typically contains a technology-independent optimization phase followed by technology mapping.  
Technology-independent optimization applies various transformations to minimize design costs represented by technology-independent metrics such as graph node count and logic level. This step focuses on optimizing the logic structure of the circuit without considering specific gate-level implementations.\looseness=-1

One of the commonly used modern logic synthesis frameworks is \texttt{ABC}~\cite{mishchenkoabc}, which is based on the And-Inverter Graph (AIG) representation of logic circuits. \texttt{ABC} is equipped with various algorithms (for example, \texttt{rewrite}, \texttt{refactor}, and \texttt{resubstitute}~\cite{resub,AIG,BooleanRewriting}) to reduce the node count or logic level in an AIG.
These algorithms follow a common design concept: they attempt incremental changes to a fraction of the graph (e.g., a k-feasible cut) and each chooses the logic form with the most cost saving.  
This idea has achieved great success in multi-level logic synthesis. However, there are two major challenges that limit the optimality of the synthesized circuit. 
First, the local decisions of logic rewriting are greedy in nature as they do not account for the influence of other optimization steps. Consequently, a sequence of local optimizations may lead to a local minima and result in a tremendous loss of optimization opportunities.
Second, the technology-independent cost metrics, such as AIG node count, may not always reflect post-mapping QoR. 
\Cref{fig:netlist-after-rewrite} gives such an example. It compares the gate-level netlists after applying different optimizations. The original logic form contains 20 AIG nodes. While \texttt{rewrite} reduces the node count to 17, the area after technology mapping actually goes up.


Motivated by the limitations of existing works, we introduce \deftextEsyn, a novel logic optimization method that utilizes equivalence graphs (e-graphs) in Boolean logic rewriting. E-graph is a data structure that preserves equivalence during rewriting-based optimizations. Thanks to the efficient and concise representation, it is possible to keep a large set of equivalent forms of the same logic. 
We can defer candidate selection until the completion of rewriting and therefore, can take a more global view when choosing from equivalent candidates.
This unique feature of e-graph 
makes it easier to explore more logic forms in the search for an optimal gate-level implementation.
Meanwhile, \deftextEsyn allows the use of customizable cost functions such as a machine learning model that directs optimization towards a specific technology-dependent target.
%
%
The last two netlists in Figure~\ref{fig:netlist-after-rewrite} show the results of delay-oriented and area-oriented \deftextEsyn optimizations. 
In this example, \deftextEsyn does not blindly reduce the corresponding AIG nodes which may not always lead to actual area saving. Instead, it targets post-mapping QoR and obtains logic forms with a much lower delay and a comparable area consumption.\looseness=-1

Specifically, this paper makes the following contributions:\looseness=-1

\begin{itemize}
    \item  It presents a novel logic optimization framework \deftextEsyn that leverages e-graph for combinational logic rewriting. \deftextEsyn maintains equivalent classes of the logic specification and can 
    explore a wider range of equivalent logic forms to search for an optimal design.

    \item This is the first work that extends the application of e-graph to Boolean logic optimization at the gate level, 
    while previous studies concentrated on applying e-graph optimizations at higher levels (such as RTL or high-level synthesis).

    \item We propose a technology-aware logic optimization method using cost models obtained from XGBoost regression. 
    The integration of the machine learning bridges the gap between technology-independent logic cost and post-mapping QoR. \looseness=-1

    \item For e-graph processing, we bring up a \textit{pool extraction} method to accommodate customizable cost functions in e-graph extraction, which scales better than the time-consuming integer linear programming method in the prior work and lifts the linear and monotone restrictions on the cost function. It also outperforms the default extraction heuristics by 21\% and 10\% in delay and area, respectively. 
    \item  We conduct a comprehensive evaluation based on post-mapping QoR on benchmark circuits from EPFL~\cite{amaru2015epfl}, LGSynth~\cite{yang1989logic,yang1991logic}, ITC~\cite{corno2000rt}, ISCAS~\cite{brglez1985neutral} etc.
    Experiments show \deftextEsyn improves 15.29\% in delay for delay-oriented optimization compared to the state-of-the-art AIG-based logic optimizations in \texttt{ABC}, and achieves 6.42\% area saving in area-oriented synthesis.
\end{itemize}

The rest of the paper is structured as follows: in the next section, we discuss the related work of logic synthesis and the applications of e-graph in optimization.
~\Cref{sec:method} presents the workflow of \deftextEsyn, followed by experiment results in ~\Cref{sec:experiment}. 
Finally, the paper concludes with ~\Cref{sec:conclusion}.

%% file: sections/related.tex
\section{Related Work}
\label{sec:related}

\subsection{Multi-level logic synthesis}
Modern multi-level logic optimization techniques work on common technology-independent logic representations, such as And-Inverter-Graphs (AIGs)~\cite{AIG}, Majority-Inverter-Graphs (MIGs)~\cite{MIG}, and XOR-based representations~\cite{XIG, XAIG}. Optimizations are centered around reducing certain metrics defined upon these representations, such as the graph node count or the longest path. 
For example, in the commonly-used logic synthesis tool \texttt{ABC}~\cite{mishchenkoabc}, the \texttt{rewrite} operation traverses the graph to find opportunities to replace a k-feasible cut with a logic-equivalent form given that the replacement brings about the most node count decrease~\cite{AIG}. 
Other AIG rewriting techniques~\cite{AIG,resub, dchoice}, such as \texttt{resubstitute}, \texttt{refactor} and \texttt{balance}, follow a similar fashion 
that they work on the subgraphs and apply local changes each time.
However, it is generally a hard question of how to precisely evaluate the different choices of local rewriting at each step.
E-graph-based optimization in \deftextEsyn differs from the prior works as it keeps the equivalent classes in the graph. There is a separate \textit{extraction} step after rewriting to pick the best logic form for the whole graph. At this step, the selection decision may refer to a more general cost model that predicts the actual delay or area cost with features from the whole graph.
\subsection{Optimization using e-graphs}
An e-graph, or equivalence graph, is a data structure that compactly represents a large number of equivalence relations.
It has been used inside automated theorem provers, for example, Z3~\cite{de2008z3}. 
Recent works have demonstrated the power of e-graph in rewriting-driven optimizations.
Coward \textit{et al.} proposed a datapath optimization approach that represents designs as data-flow graphs and 
leverages e-graphs and equality saturation techniques for efficient design space exploration at the register-transfer-level~\cite{coward2022automatic}. 
The IMpress framework~\cite{ustun2022impress} employed e-graphs to tackle the implementation problem of large integer multiplication in high-level synthesis (HLS), where e-graph helps to explore the possible ways to to decompose multipliers corresponding to different hardware implementations~\cite{ustun2023equality}. Applying e-graph in optimization solves the phase ordering problem as the ordering of transformations in e-graph is less of a concern.
Besides optimization, in the work~\cite{coward2023datapath}, the authors proposed an equivalence checking method that utilizes the e-graph to rewrite RTL for complex datapaths. E-graph in verification transcends the traditional bit-level logic reasoning and helps to prove equivalence at a higher level.
These aforementioned prior works were based by \texttt{egg}~\cite{10.1145/3434304}, a fast and flexible open-source library, which provides built-in functionality for efficient e-graph manipulation and
an extendable interface to incorporate domain-specific rewriting rules. 

Compared to the prior works, this paper extends e-graph to bit-level optimization with Boolean algebra rules for rewriting. To effectively process the large-scale logic formulas in digital circuits, we devise efficient format converters to integrate the \texttt{egg} library into the logic synthesis flow and put forward a fast and objective-aware pool extraction method to accommodate nonlinear technology-aware cost models in logic optimization.\looseness=-1

%% file: sections/our-method.tex
 \begin{figure*}[htbp]
  \centering
  \includegraphics[width=0.9\textwidth]{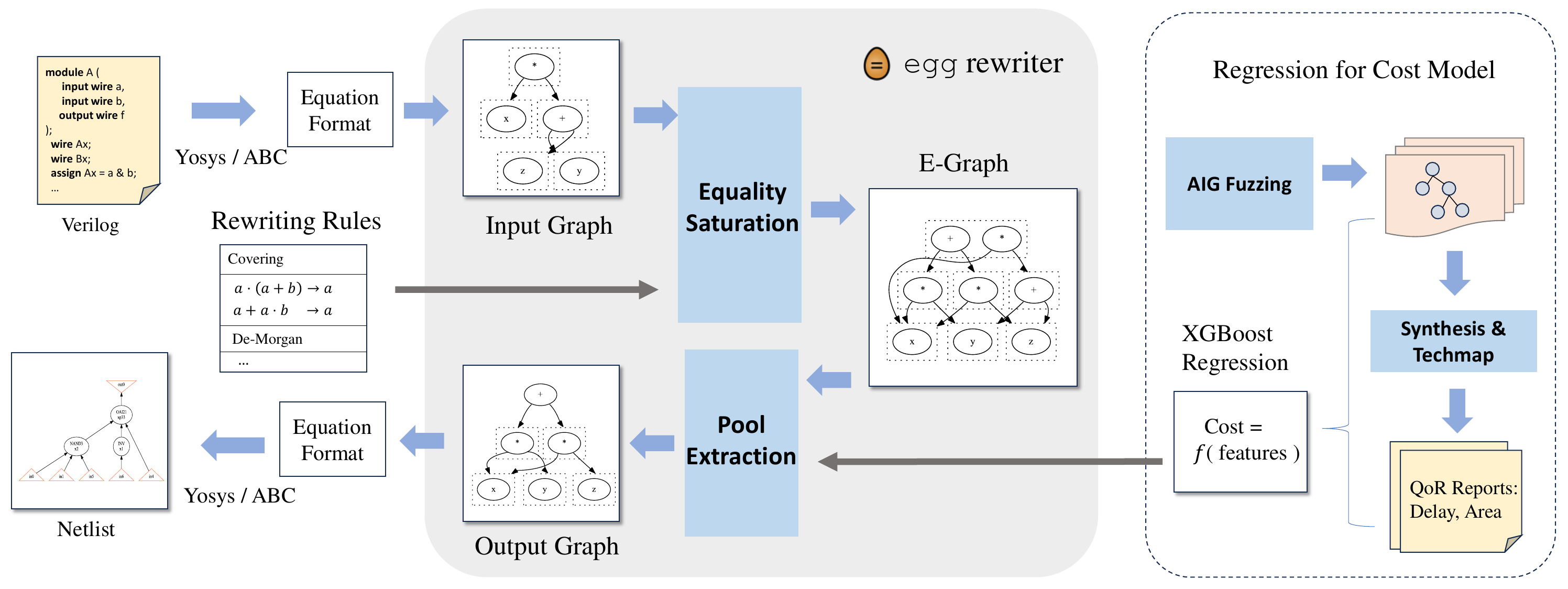}
  \caption{The framework of \deftextEsyn}
  \label{fig:Flow-chart}
\end{figure*}

\begin{figure}
    \centering
    \includegraphics[width=0.3\textwidth]{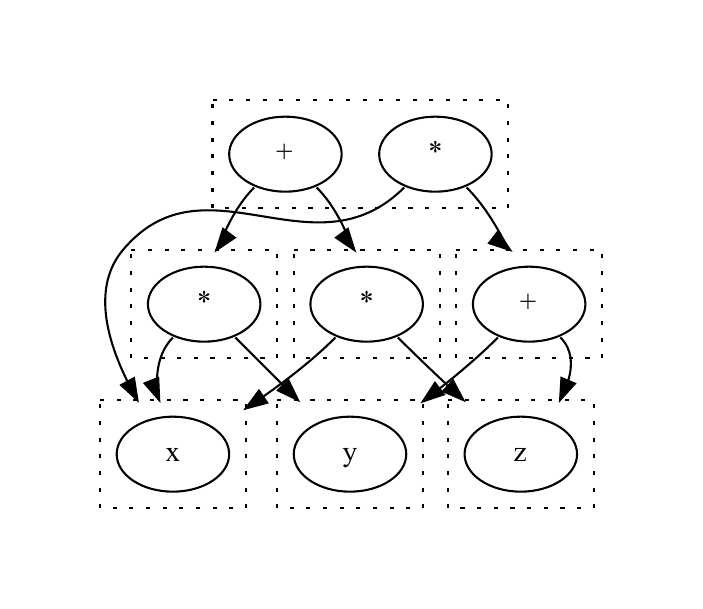}
    \caption{An example of e-graph for logic function $xy+xz$ where we use ``$*$'' for \texttt{AND} and ``$+$'' for \texttt{OR}.}
    \label{fig:egraph-example}
\end{figure}

\section{E-Graph Rewriting for Logic Synthesis}\label{sec:method}

In this section, we introduce how e-graph rewriting is used to optimize Boolean logic. \Cref{fig:Flow-chart} shows the overall workflow of our proposed method. 
\deftextEsyn is built upon the classic Yosys/ABC logic synthesis flow with additional e-graph optimization steps to allow a wider exploration for optimal design forms.


\begin{table}
  \centering
  \caption{Rewriting Rules in \deftextEsyn} 
  \label{tab:rewrites}
  \resizebox{0.87\columnwidth}{!}{
  \begin{tabular}{>{\raggedright\arraybackslash}m{2cm}>{\raggedright\arraybackslash}m{6cm}}
    \toprule
    \textbf{Class} & \textbf{Boolean Rewriting Rules} \\
    \midrule
    \ & $a * 1 \Rightarrow  a$\\
    \multirow{3}{*}{\textbf{Complements}} & $ a * 0 \Rightarrow 0$\\
    \ & $a + 1 \Rightarrow  1$
     \\
     \ & $(\neg a) * a \Rightarrow  0$ \\
     \ & $(\neg a) + a \Rightarrow 1$ \\
    \midrule    
    \multirow{2}{*}{\textbf{Covering}} &  $a * (a + b) \Rightarrow a $ \\
    \ &  $a + (a * b) \Rightarrow a$
    \\
    \midrule    
    \multirow{2}{*}{\textbf{Combining}} & 
    $((a * b) + (a * \neg b )) \Rightarrow a$
     \\
    \ & 
    $((a + b) * (a + \neg b)) \Rightarrow a$
    \\
    \midrule    
    
    \multirow{2}{*}{\textbf{Idempotency}} & 
$a * a \Rightarrow a$
 \\

\ & $a + a \Rightarrow a$
\\
\midrule    
    \multirow{2}{*}{\textbf{Commutativity}} & 
    $a * b \Leftrightarrow b * a$
 \\

\ & $a + b \Leftrightarrow b + a$
\\
\midrule    
    \multirow{2}{*}{\textbf{Associativity}} & 
    $(a * b) * c  \Leftrightarrow a * (b * c)$
 \\
\ & 
$(a + b) + c  \Leftrightarrow a + (b + c)$
\\
\midrule    
    \multirow{3}{*}{\textbf{Distributivity}} & 
    $a * (b + c) \Rightarrow a * b + a * c $
 \\
\ & $(a + b) * (a + c) \Rightarrow a + (b * c)$
    $(a * b) + (a * c) \Rightarrow a * (b + c)$
\\
\midrule    
    \multirow{2}{*}{\textbf{Consensus}} & 
    $(a * b) + ((\neg a) * c) + b * c  \Rightarrow (a * b) + (\neg a) * c$
 \\

\ & $((a + b) * ((\neg a) + c)) * (b + c) \Rightarrow (a + b) * ((\neg a) + c)$
\\
\midrule    
        \multirow{2}{*}{\textbf{De-Morgan}} & $\neg(a *b ) \Rightarrow \neg a + \neg b$ 
 \\
\ & $\neg(a + b) \Rightarrow (\neg a) * (\neg b)$
\\
\bottomrule
    
  \end{tabular}
  }
\end{table}

\subsection{Rewriting using E-graph}

In \deftextEsyn, we take advantage of e-graph to efficiently represent the Boolean logic formulas that are equivalent to the given function specification.
In an e-graph, equivalent Boolean functions are clustered into equivalent classes (e-classes). For example, \Cref{fig:egraph-example} demonstrates 
an e-graph for the logic function $xy+xz$, where each dotted box is an e-class. Nodes are maximally reused in the parent functions to avoid the repetition of equivalent nodes. 
Thanks to the compact representation of nested e-classes, a number of e-graph nodes can potentially represent exponentially many equivalent forms.
E-graph does not restrict the set of operators to use. Though it is possible to use only AND-gates and inverters to mimic AIG rewriting, we decide to loosen the requirement on the operators and allow free use of \texttt{AND}, \texttt{OR} and \texttt{NOT} to match the input logic function specification.

The construction of an e-graph utilizes the equality saturation technique~\cite{EQsaturation}, which has been well-established in the formal methods and compiler research communities.
 Equality saturation applies a series of rule-based transformations to generate equivalent representations of the given function. In \deftextEsyn, we use the laws of Boolean algebra shown in ~\Cref{tab:rewrites} as the rewriting rules. 
``$\Leftrightarrow$'' in the table means the rewriting rule is applied in both directions, while for a few rules that serve as purely a simplification, we only apply it from left to right, as indicated by ``$\Rightarrow$''.
Unlike typical logic rewriting techniques (e.g., DAG-aware AIG rewriting) that heuristically select among equivalent representations locally at the rewriting step, \deftextEsyn keeps those equivalent representations in the graph and takes a separate extraction step to select the best candidate after the completion of rewriting.

\subsection{Extraction}\label{sec:pool-extraction}

The extraction step traverses the e-graph to select an optimal logic form for each node. For each e-class with more than one element, it needs to choose based on a certain cost model. The selection finally returns an abstract syntax tree (AST) from the graph.
Prior works have used the depth or size of the AST
as the cost function. However, these cost functions may not well reflect the actual area or delay cost of the circuit netlist after technology mapping. Therefore, to better account for the actual QoR in the extraction step, we make use of regression to obtain a more technology-aware cost model.

\subsubsection{Regression}\label{sec:regression}
Inspired by the regression model used in RTL-stage QoR prediction~\cite{how-good-is-your-rtl}, we can employ an XGBoost model to fit the area and delay cost from the AST of a Boolean expression. The features used in the regression are listed as follows:

\begin{itemize}
    \item \textbf{Boolean operator count.} For each type of Boolean operator, we count their occurrence in the AST. 
    \item \textbf{AST node count.}  The number of nodes in an AST could reflect the overall scale of the logic circuit and therefore could be one feature in the regression.
    \item \textbf{AST depth.} AST depth serves a similar purpose as the logic level in logic synthesis, which could correlate with the length of paths and therefore, the delay of the circuit.
    \item \textbf{Graph density and edge sum.} AST can also be regarded as a graph and we utilize the commonly-used graph features. In our regression model, we consider the graph density and edge sum as two features. Roughly speaking, these two features indicate the abundance of edges in a graph. 
\end{itemize}

To generate the training data for regression, we use \texttt{aigfuzz} in the AIG library\footnote{https://github.com/arminbiere/aiger} to create a dataset of 50000 random-sized combinational logic circuits with an average logic level of 234 and an average size of 6305 AIG nodes. These circuits are then converted into the equation format and transformed into e-graphs for feature extraction. We further run technology mapping and technology-dependent optimization on these circuits using \texttt{ABC}. The reported delay and area at this step are used as the training labels. We use two separate XGBoost regression models to predict area and delay, respectively. Each contains 200 estimators and has a maximum depth of 5. 
The delay prediction achieves an R-value of 0.78,  and the R-value of area prediction is 0.76. 
To integrate the XGBoost model with \texttt{egg} library for e-graph extraction, we use the \texttt{Rust} binding of the XGBoost library. \looseness=-1


\subsubsection{Extraction with Technology-Aware Cost}
The existing extraction  methods in the prior works either (1) depend on local heuristic decisions per e-node or (2) require solving an integer linear programming (ILP) problem.
For extractor (1), the local heuristics primarily rely on local cost functions such as the size or depth of a graph, which may significantly diverge from the actual cost. 
While for existing extractor (2), though it is global as it formulates extraction into an ILP problem to solve, it limits the form of the cost function to be linear and monotonically increasing from leaf nodes to parent nodes.
In order to best fit the technology-dependent cost, the cost model might not be linear and monotone. Besides, it is generally hard to scale ILP up for large-scale Boolean logic functions in practical digital circuits. Therefore, in this application of e-graph, we are looking for a fast and flexible extraction method. 

With the consideration of the pros and cons of the two existing methods, we bring up the novel \textit{pool extraction method} in \deftextEsyn to achieve efficient logic function extraction using technology-aware cost models. As its name suggests, pool extraction first collects a pool of candidates from an e-graph using a combination of heuristics, and then it evaluates each candidate using the given cost model to pick the best one in the pool.
Pool extraction combines the above two existing extraction methods and serves as a trade-off between efficiency and performance. It also removes the limitations on the cost functions.

Specifically, the candidate pool consists of one AST with the fewest number of nodes, one with the least tree depth, and  candidates from sampling in the e-graph. 
The sampling process traverses the e-classes in the e-graph with two strategies: (a) randomly select an e-node only from those with the same least local cost, or (b) select an e-node that has a sub-optimal local cost with a probability. 
The first strategy is different from the default extractor (1) as we introduce randomness, whereas the default extractor will always choose the first candidate among those with the same least cost. The second strategy occasionally explores a choice with sub-optimal local cost and therefore, can potentially find a form with better global cost.
Here, the local cost is one of the AST depth, AST size, or the weighted sum of operators (we assign  a lower weight to  \texttt{NOT} than \texttt{AND}, \texttt{OR}).
We empirically set the probability of sub-optimal exploration to $0.2$ and the ratio of candidates sampled from these two strategies as 1:3.

After we obtain a pool of ASTs following the above strategies, we will evaluate each candidate using the technology-aware cost model obtained in \Cref{sec:regression}. The one with the lowest cost will be selected as the optimal logic form for the given function specification.


\subsection{Integration with the Existing Synthesis Flow}

E-graph rewriting is not exclusive to the existing logic synthesis flow. Instead, it is designed as an enhancement to be inserted at the beginning of a logic optimization flow to improve the final QoR. With the command \texttt{write\_eqn} from ABC, a combinational logic circuit in the AIG synthesis flow can be written into the equation format, which contains Boolean \texttt{AND}, \texttt{OR}, \texttt{NOT} operators, intermediate variables and nested parentheses. The equation format specification is then transformed into nested S-expressions in \texttt{Common Lisp}, 
which is the input format for \texttt{egg},
the library for e-graph rewriting and equality saturation. The output AST from \texttt{egg} is then converted back to the equation format and can be later processed by the traditional logic optimization flow. We also check the result using combinational equivalence checking 
to ensure correct implementation of logic rewriting in e-graph.
As for the implementation, we design high-performance parsers 
in \texttt{Rust} for the above format conversion, which supports parallel execution to minimize the influence on the synthesis time.

%% file: sections/experiment.tex
\section{Experiment Result and Discussion}
\label{sec:experiment}

\subsection{Experiment Setup}

E-graph rewriting and extraction are implemented in \texttt{Rust} to interface with the \texttt{egg} library. 
All experiments are performed on a Ubuntu 20.04.4 LTS server equipped with Intel Xeon Platinum 8375C processors, an NVIDIA 3090 GPU, and 128GB memory.
The test circuits are primarily arithmetic blocks, including those from LGSynth~\cite{yang1989logic,yang1991logic}, ITC~\cite{corno2000rt}, EPFL~\cite{amaru2015epfl} and ISCAS85~\cite{brglez1985neutral} benchmark suites. 
We also use \texttt{genmul}~\cite{mahzoon2021genmul} to generate two multipliers (3x3 and 5x5).
Additionally, we take an open-source divider from OpenCores\footnote{https://opencores.org/projects/verilog\_fixed\_point\_math\_library}. 
The arithmetic circuits cover a wide range of complexity levels, ranging from simple adders to complex multipliers, and therefore, provide a diverse set of benchmarks that can test different aspects of the logic optimization algorithm.
The setup for equality saturation runtime limit is 300 seconds and the e-node limit is 2500000 nodes. 
Throughout these experiments, we use the ASAP 7nm technology library~\cite{clark2016asap7}.

\begin{figure}[tbp]  
    \centering
    \begin{minipage}{0.49\linewidth}
        \centering
        \includegraphics[width=\linewidth]{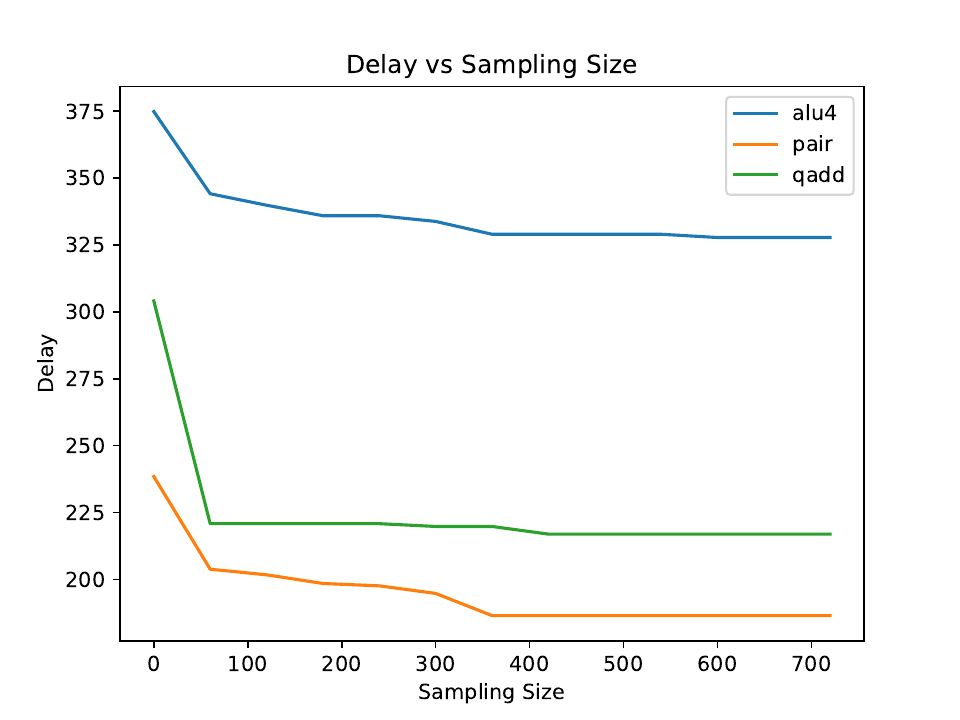}
    \end{minipage}
    \hfill  
    \begin{minipage}{0.49\linewidth}
        \centering
        \includegraphics[width=\linewidth]{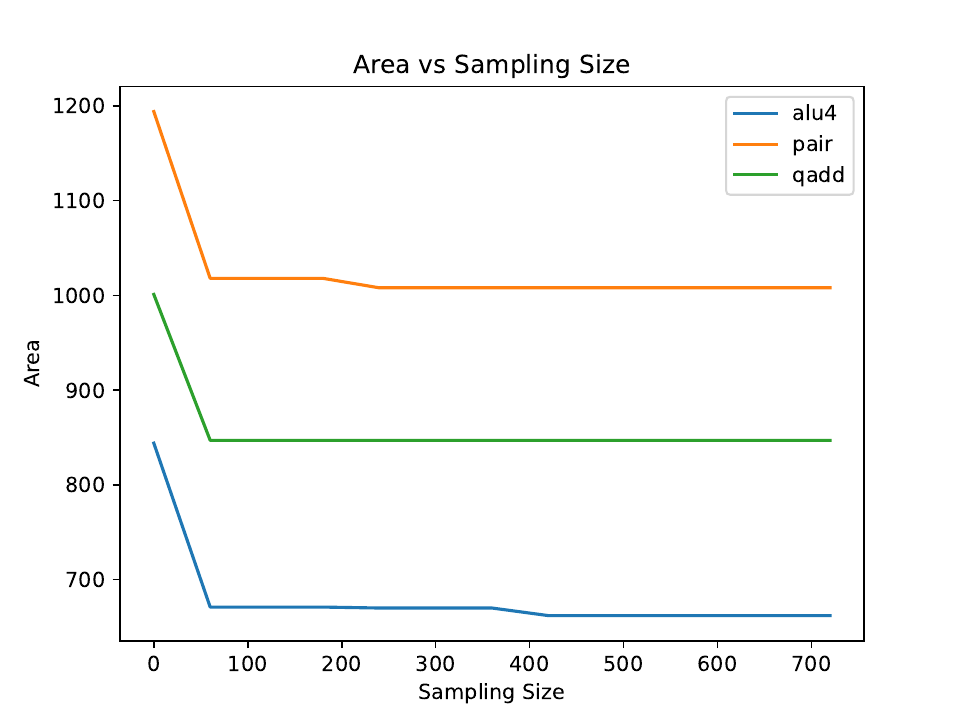}
    \end{minipage}
    \caption{Sampling size vs. minimum delay and area} 
    \label{fig:pool-size}
\end{figure}

\input{sections/qor_exp_table}

\subsection{Effectiveness of the Pool Extraction Method}
For fast and efficient extraction using a technology-aware cost model, we design the pool extraction method in \Cref{sec:pool-extraction}.
To pick a proper sampling size for our pool extraction method, we conduct experiments to measure the influence of sampling size to the QoR.
We incrementally sample more candidates on the e-graphs of test circuits while measuring the technology-dependent costs using the following command:
\textit{\textbf{ strash; dch -f; map; topo; upsize; dnsize; stime}}. 
We take record of the best area and delay among all candidates in the pool under different sampling pool sizes, and plot their relation with the sampling size in \Cref{fig:pool-size}. For QoR under a small sampling size ($<100$), the best in the pool may vary from run to run due to the randomness in sampling, and it becomes more deterministic for larger sampling sizes.
It can be seen that there is a diminishing return from expanding the sampling pool. A pool size of over 100 would suffice in most cases. 

\begin{figure}[ht]
    \centering
    \includegraphics[width=0.9\columnwidth]{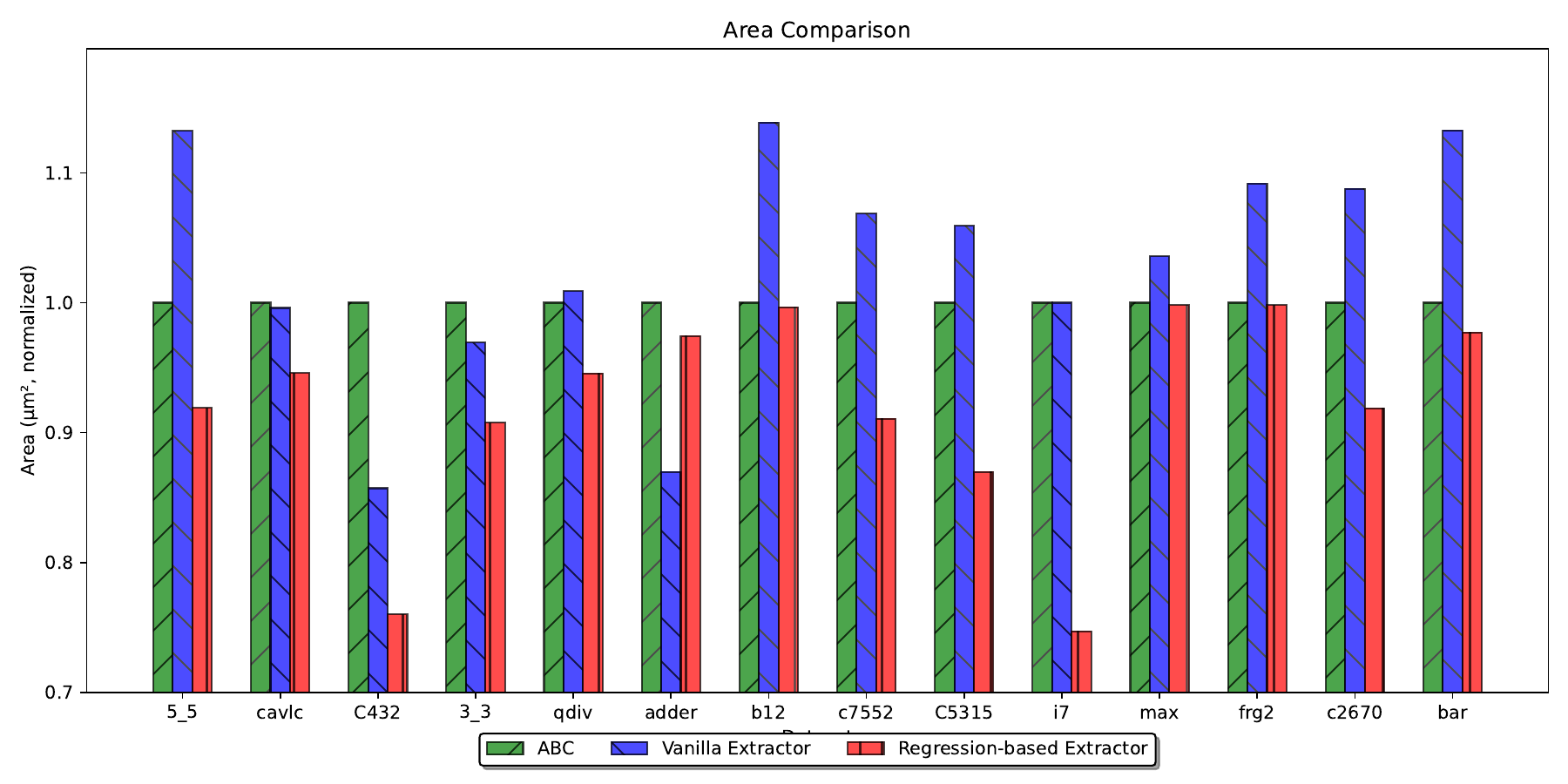}
    \par\bigskip 
    \includegraphics[width=0.9\columnwidth]{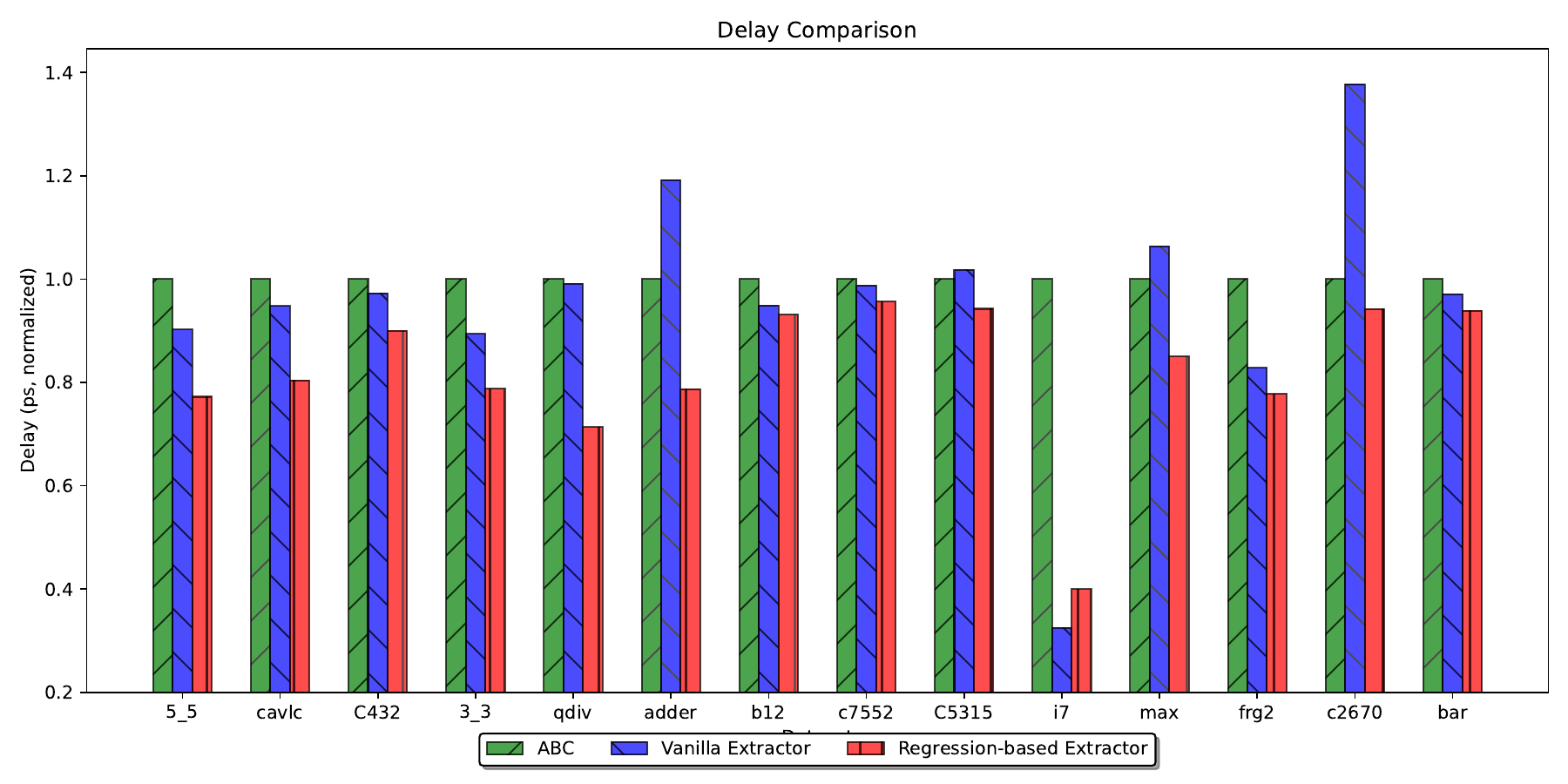}
    \caption{Comparison of e-graph optimization using the vanilla extractor vs. our pool extraction. Results are normalized by QoR from the baseline \texttt{ABC} flow.}
    \label{fig:both-charts}
\end{figure}

We also compare our proposed pool extraction method to the vanilla extraction methods in the \texttt{egg} library.
Because we use the XGBoost regression model as the cost function, which is not linear, it is not feasible to compare with the ILP method.
The other built-in extractor for comparison is the default greedy extractor that takes either AST size or AST depth as the cost function.
For delay comparison, we use AST depth as the cost function for the vanilla \texttt{egg} extractor and the delay regression model for the pool extraction method. For area comparison, we use AST size in vanilla extractor and the area regression model for pool extraction.
We assess the final delay and area cost by the same \texttt{ABC} flow above.
%
%
The results are plotted in \Cref{fig:both-charts}, where delay and area metrics are normalized by the QoR when no e-graph rewriting is used, as indicated by the legend ABC in the figure.
Our findings suggest that e-graph rewriting with the vanilla extractor may not always improve delay or area. It is necessary to use pool extraction with technology-aware cost models.
 Compared to the vanilla extractor, the pool extraction method achieves up to 25\% area saving (avg. 10\%), and up to 34\% of reduction on delay (avg. 21\%). Overall, it improves from the baseline \texttt{ABC} flow by 6\% and 18\% in area and delay, respectively.

\begin{figure}[tbp]  
    \centering
    \begin{minipage}{0.49\linewidth}
        \centering
        \includegraphics[width=\linewidth]{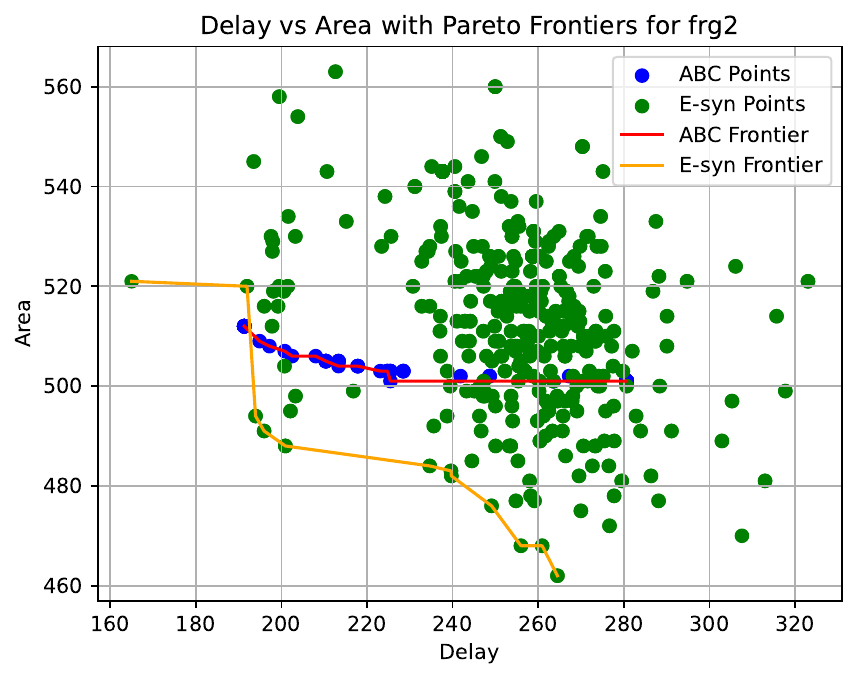}
    \end{minipage}
    \hfill  
    \begin{minipage}{0.49\linewidth}
        \centering
        \includegraphics[width=\linewidth]{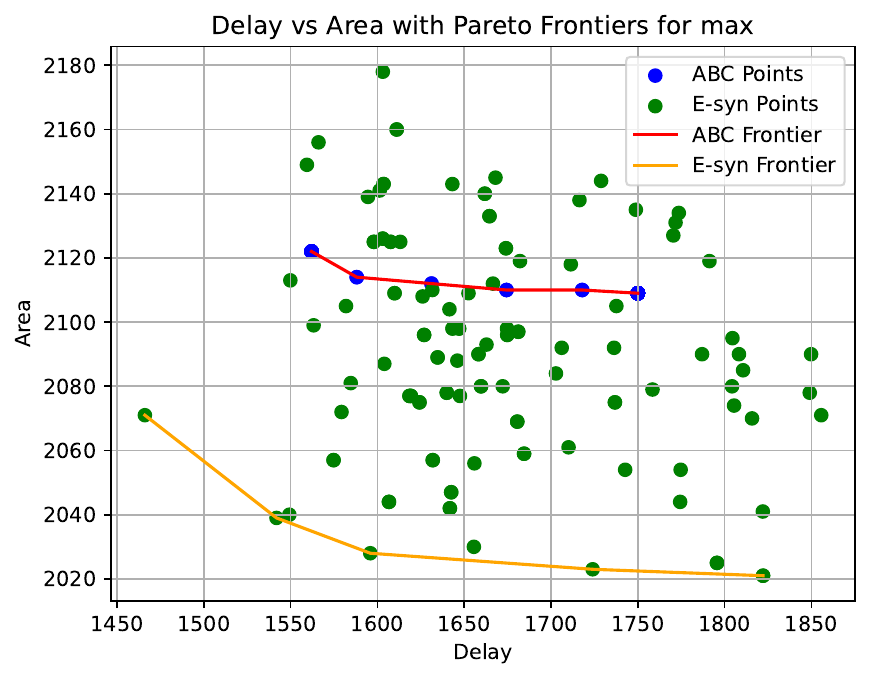}
    \end{minipage}
    \caption{Comparison of design space and Pareto-frontier of \deftextEsyn and AIG rewriting} 
    \label{fig:pareto-frontier}
\end{figure}

\subsection{Comparison on the Design Space}

To further validate our claim that e-graph logic rewriting can reach a wider design space, we conduct experiments to compare  circuits produced by \deftextEsyn to those resulted from a commonly-used logic synthesis flow in \texttt{ABC}.
The default \text{ABC} synthesis flow available as the \textit{\textbf{abc}} command in Yosys is as follows: \textit{\textbf{ strash; ifraig; scorr; dc2; dretime; retime -o -D \{delay\}; strash; \&get -n; \&dch -f; \&nf -D \{delay\}; \&put; buffer; upsize -D \{delay\}; dnsize -D \{delay\}; stime}}.  It contains a delay constraint parameter that can be set to control the delay-area trade-off in synthesis.
In addition to the various AIG rewriting operations performed by the \textit{\textbf{dc2}} command,
this flow also makes use of techniques like structural hashing, correlated signal reduction to further optimize the given logic.  Moreover, the \textit{\textbf{\&dch}} command combines different networks seen during technology-independent synthesis into a single network with choices. This is to help the subsequent technology mapping to better choose the logic structure with the optimal costs.
We argue that this synthesis flow in comparison is already a relatively powerful one. 
We tune the target delay in this script to generate designs with different area-delay trade-offs and plot the QoR on the delay-area plane. These make up the \texttt{ABC} design points for comparison.
As for the \deftextEsyn flow, we plot the QoR of all candidates in the pool for comparison.

\Cref{fig:pareto-frontier} shows the design points for a medium-size and a large-size test circuit.  In general, the design points from \deftextEsyn span a wider range in the delay-area plane. In both designs, the frontier of \deftextEsyn completely dominates. 
%
We also compare QoR for the test circuits under different constraints (delay-oriented, area-oriented, and balanced). The experiment results are shown in \Cref{tab:all-results}. 
Though AIG-based rewriting has been extensively optimized towards area, our method can further reduce the area cost by a margin of 6.42\%, if the users are willing to sacrifice more delay. On the side of delay-oriented optimization, \deftextEsyn achieves a delay reduction on almost all designs, averaging 15.29\%.
For the delay-area-balanced optimization target, our method outperforms in both delay and area. Overall, the comparison over three regions in the design space indicates the Pareto-frontier of \deftextEsyn dominates the above \texttt{ABC} synthesis flow.

Regarding runtime,  \deftextEsyn flow  takes 80 seconds on average to run  on a test circuit, in addition to the 
300-seconds time-limit for equality saturation, which  may be lowered to trade quality for time.

%% file: sections/qor_exp_table.tex
\begin{table*}
    \caption{QoR of \deftextEsyn and \texttt{ABC} synthesis flow under different constraints}
    \centering
    \resizebox{\linewidth}{!}{%
    \begin{tabular}{l|rr|rr|rr|rr|rr|rc}
    \hline
    \multicolumn{1}{c|}{\multirow{2}{*}{Circuit}} &
    \multicolumn{2}{c|}{ABC delay-oriented} & 
    \multicolumn{2}{c|}{\deftextEsyn delay-oriented} & 
    \multicolumn{2}{c|}{ABC area-oriented} & 
    \multicolumn{2}{c|}{\deftextEsyn area-oriented} & 
    \multicolumn{2}{c|}{ABC balanced} & 
    \multicolumn{2}{c}{\deftextEsyn balanced} 
    
    \\ \cline{2-13} 
     &
    \multicolumn{1}{c}{Area ($\mu m^2$)} & 
    \multicolumn{1}{c|}{Delay ($ps$)} & 
    \multicolumn{1}{c}{Area ($\mu m^2$)} & 
    \multicolumn{1}{c|}{Delay ($ps$)} & 
    \multicolumn{1}{c}{Area ($\mu m^2$)} & 
    \multicolumn{1}{c|}{Delay ($ps$)} &
    \multicolumn{1}{c}{Area ($\mu m^2$)} & 
    \multicolumn{1}{c|}{Delay ($ps$)} &
    \multicolumn{1}{c}{Area ($\mu m^2$)} & 
    \multicolumn{1}{c|}{Delay ($ps$)} & 
    \multicolumn{1}{c}{Area ($\mu m^2$)} & 
    \multicolumn{1}{c}{Delay ($ps$)}
 \\
    \hline        
adder~(EPFL)      & 988  & 2172.78 & 988           & \textbf{2168.92} & 981  & 2182.02 & 981            & \textbf{2178.23}  & 983  & 2173.38 & 988           & \textbf{2169.75}       \\
bar~(EPFL)        & 2262 & 197.82  & 2266          & 198.21           & 2235 & 234.2   & \textbf{2141}  & 307.91            & 2238 & 218.63  & \textbf{2201} & \textbf{206.8}         \\
max~(EPFL)        & 2122 & 1562.07 & \textbf{2071} & \textbf{1466.06} & 2109 & 1750.01 & \textbf{2021}  & 1822.13           & 2112 & 1631.15 & \textbf{2105} & \textbf{1581.91}       \\
cavlc~(EPFL)      & 452  & 150.2   & 466           & \textbf{129.12}  & 434  & 186.19  & \textbf{415}   & 211.37            & 441  & 151.49  & 442           & \textbf{149.67}        \\
3\_3~(genmul)     & 41   & 146.44  & 44            & \textbf{113.28}  & 37   & 165.24  & \textbf{33}    & 181.14            & 40   & 146.95  & \textbf{37}   & \textbf{117.17}        \\
5\_5~(genmul)     & 144  & 424.64  & \textbf{132}  & \textbf{329.64}  & 130  & 466.33  & \textbf{116}   & \textbf{402.48}   & 135  & 437.86  & \textbf{120}  & \textbf{422.85}        \\
qdiv~(opencore)   & 1123 & 747.75  & 1280          & \textbf{465.18}  & 1101 & 812.67  & \textbf{1089}  & \textbf{709.38}   & 1103 & 755.68  & \textbf{1102} & \textbf{648.15}        \\
C5315~(LGSynth91) & 1075 & 351.03  & \textbf{1043} & \textbf{314.19}  & 1058 & 384.14  & \textbf{1012}  & 401.84            & 1062 & 367.51  & \textbf{1050} & \textbf{347.12}        \\
i7~(LGSynth91)    & 477  & 96.69   & \textbf{347}  & \textbf{93.98}   & 468  & 162.32  & \textbf{321}   & 180.81            & 473  & 103.39  & \textbf{345}  & \textbf{99.85}         \\
c7552~(ISCAS85)   & 1191 & 465.85  & 1298          & \textbf{299.55}  & 1176 & 587.9   & \textbf{1175}  & \textbf{470.56}   & 1185 & 482.12  & \textbf{1182} & \textbf{459.39}        \\
c2670~(ISCAS85)   & 536  & 240.84  & 537           & \textbf{200.65}  & 494  & 299.45  & \textbf{481}   & 304.84            & 522  & 256.67  & \textbf{516}  & \textbf{219.55}        \\
frg2~(LGSynth89)  & 512  & 191.34  & 521           & \textbf{165.04}  & 501  & 280.64  & \textbf{470}   & 307.57            & 505  & 210.4   & \textbf{488}  & \textbf{200.95}        \\
C432~(LGSynth89)  & 98   & 372.69  & 112           & \textbf{335.19}  & 95   & 451.72  & \textbf{91}    & \textbf{368.79}   & 96   & 396.51  & \textbf{94}   & \textbf{363.48}        \\
b12~(ITC99)       & 770  & 244.33  & 776           & \textbf{219.74 } & 736  & 303.24  & \textbf{734}   & \textbf{299.32}   & 750  & 257.07  & \textbf{747}  & \textbf{249.71}        \\

    \hline   
GEOMEAN      & 540.10 & 342.93 & 540.90 & \textbf{290.51} & 520.49 & 414.37 & \textbf{487.10} & 416.50 & 529.97 &358.34 &\textbf{507.39}  &\textbf{334.28}  \\ \hline
Improvements &             &             &             & \textbf{15.29\%} &             &             & \textbf{6.42\%} &             &  &  &\textbf{4.26\%} &\textbf{6.71\%}  \\ \hline
\end{tabular}}
\label{tab:all-results}

\end{table*}

%% file: sections/conclusion.tex
\section{Conclusion}
\label{sec:conclusion}

This paper proposes using e-graph rewriting in logic synthesis. It extends e-graph optimization to the bit level. E-graph-based optimization explores a wider range of logic forms than local logic rewriting. It can also factor in technology-aware costs to better target delay or area optimization in logic synthesis.


%% file: sections/ack.tex
\section*{Acknowledgments}
This work is supported in part by Guangzhou Municipal Science and Technology Project (Municipal Key Laboratory Construction Project, Grant No.2023A03J0013) and by National Natural Science Foundation of China (Grant No. 62304194).